 \documentclass[final,5p,times,twocolumn,authoryear]{elsarticle}

\usepackage{amssymb}
\usepackage{amsmath}
\usepackage{hyperref}
\usepackage{booktabs}
\usepackage{newtxtext}

\usepackage{lineno}
\usepackage{comment}
\usepackage{caption}
\usepackage{subcaption}
\usepackage{multirow}
\usepackage{natbib}
\setcitestyle{numbers}

\journal{Nuclear Physics A}

\begin{document}

\begin{frontmatter}

\title{Signal shape studies and rate dependence of HFO-based gas mixtures in RPC detectors}

\author[1]{\small \textbf{The RPC ECOGas@GIF++ Collaboration: }L. Quaglia} 
\author[6,3]{\small M. Abbrescia}
\author[21]{\small G. Aielli}
\author[3,8]{\small R. Aly}
\author[17]{\small M. C. Arena}
\author[11]{\small M. Barroso}
\author[4]{\small L. Benussi}
\author[4]{\small S. Bianco}
\author[16]{\small F. Bordon}
\author[7]{\small D. Boscherini}
\author[7]{\small A. Bruni}
\author[18]{\small S. Buontempo}
\author[16]{\small M. Busato}
\author[21]{\small P. Camarri}
\author[13]{\small R. Cardarelli}
\author[3]{\small L. Congedo}
\author[11]{\small D. De Jesus Damiao}
\author[3]{\small F. Debernardis}
\author[6,3]{\small M. De Serio}
\author[21]{\small A. Di Ciaccio}
\author[21]{\small L. Di Stante}
\author[14]{\small P. Dupieux}
\author[19]{\small J. Eysermans}
\author[12,1]{\small A. Ferretti}
\author[12,1]{\small M. Gagliardi}
\author[6,3]{\small G. Galati}
\author[12,1]{\small S. Garetti}
\author[16]{\small R. Guida}
\author[2,3]{\small G. Iaselli}
\author[14]{\small B. Joly}
\author[24]{\small S.A. Juks}
\author[23]{\small K.S. Lee}
\author[13]{\small B. Liberti} 
\author[22]{\small D. Lucero Ramirez}
\author[16]{\small B. Mandelli}
\author[14]{\small S.P. Manen}
\author[7]{\small L. Massa}
\author[3]{\small A. Pastore}
\author[13]{\small E. Pastori}
\author[4]{\small D. Piccolo}
\author[13]{\small L. Pizzimento}
\author[7]{\small A. Polini}
\author[13]{\small G. Proto}
\author[2,3]{\small G. Pugliese}
\author[2,3]{\small D. Ramos}
\author[16]{\small G. Rigoletti}
\author[13]{\small A. Rocchi}
\author[7]{\small M. Romano}
\author[9]{\small P. Salvini}
\author[10]{\small A. Samalan}
\author[21]{\small R. Santonico}
\author[5]{\small G. Saviano}
\author[13]{\small M. Sessa}
\author[6,3]{\small S. Simone}
\author[12,1]{\small L. Terlizzi}
\author[10,20]{\small M. Tytgat}
\author[12,1]{\small E. Vercellin}
\author[15]{\small M. Verzeroli}
\author[22]{\small N. Zaganidis}

\affiliation[1]{organization={INFN Sezione di Torino},
             addressline={Via P. Giuria 1},
             city={Torino},
             postcode={10125},
             state={Italy},
             country={}}
             
\affiliation[2]{organization={Politecnico di Bari, Dipartimento Interateneo di Fisica},
             addressline={via Amendola 173},
             city={Bari},
             postcode={70125},
             state={Italy},
             country={}}

\affiliation[3]{organization={INFN Sezione di Bari},
             addressline={Via E. Orabona 4},
             city={Bari},
             postcode={70125},
             state={Italy},
             country={}}

\affiliation[4]{organization={INFN - Laboratori Nazionali di Frascati},
             addressline={Via Enrico Fermi 54},
             city={Frascati (Roma)},
             postcode={00044},
             state={Italy},
             country={}}

\affiliation[5]{organization={Sapienza Università di Roma, Dipartimento di Ingegneria Chimica Materiali Ambiente},
             addressline={Piazzale Aldo Moro 5},
             city={Roma},
             postcode={00185},
             state={Italy},
             country={}}

\affiliation[6]{organization={Università degli studi di Bari, Dipartimento Interateneo di Fisica},
             addressline={Via Amendola 173},
             city={Bari},
             postcode={70125},
             state={Italy},
             country={}}

\affiliation[7]{organization={INFN Sezione di Bologna},
             addressline={Via C. Berti Pichat 4/2},
             city={Bologna},
             postcode={40127},
             state={Italy},
             country={}}

\affiliation[8]{organization={Helwan University},
             addressline={},
             city={Helwan, Cairo Governorate},
             postcode={4037120},
             state={Egypt},
             country={}}

\affiliation[9]{organization={INFN Sezione di Pavia},
             addressline={Via A. Bassi 6},
             city={Pavia},
             postcode={27100},
             state={Italy},
             country={}}

\affiliation[10]{organization={Ghent University, Dept. of Physics and Astronomy},
             addressline={Proeftuinstraat 86},
             city={Ghent},
             postcode={B-9000},
             state={Belgium},
             country={}} 
             
\affiliation[11]{organization={Universidade do Estado do Rio de Janeiro},
             addressline={R. São Francisco Xavier, 524},
             city={Maracanã, Rio de Janeiro - RJ},
             postcode={20550-013},
             state={Brazil},
             country={}} 

\affiliation[12]{organization={Università degli studi di Torino, Dipartimento di Fisica},
             addressline={Via P. Giuria 1},
             city={Torino},
             postcode={10125},
             state={Italy},
             country={}} 

\affiliation[13]{organization={INFN Sezione di Roma Tor Vergata},
             addressline={Via della Ricerca Scientifica 1},
             city={Roma},
             postcode={00133},
             state={Italy},
             country={}}

\affiliation[14]{organization={Clermont Université, Université Blaise Pascal, CNRS/IN2P3, Laboratoire de Physique Corpusculaire},
             addressline={BP 10448},
             city={Clermont-Ferrand},
             postcode={F-63000},
             state={France},
             country={}}

\affiliation[15]{organization={Universitè Claude Bernard Lyon I},
             addressline={43 Bd du 11 Novembre 1918},
             city={Villeurbanne},
             postcode={69100},
             state={France},
             country={}}

 \affiliation[16]{organization={CERN},
             addressline={Espl. des Particules 1},
             city={Meyrin},
             postcode={1211},
             state={Switzerland},
             country={}}   
             
\affiliation[17]{organization={Università degli studi di Pavia},
             addressline={Corso Strada Nuova 65},
             city={Pavia},
             postcode={27100},
             state={Italy},
             country={}} 

\affiliation[18]{organization={INFN Sezione di Napoli},
             addressline={Complesso universitario di Monte S. Angelo ed. 6 Via Cintia},
             city={Napoli},
             postcode={80126},
             state={Italy},
             country={}}

\affiliation[19]{organization={Massachusetts Institute of Technology},
             addressline={77 Massachusetts Ave},
             city={Cambridge, MA},
             postcode={02139},
             state={USA},
             country={}}

\affiliation[20]{organization={Vrije Universiteit Brussel (VUB-ELEM), Dept. of Physics},
             addressline={Pleinlaan 2},
             city={Brussels},
             postcode={1050},
             state={Belgium},
             country={}}

\affiliation[21]{organization={Università degli studi di Roma Tor Vergata, Dipartimento di Fisica},
             addressline={Via della Ricerca Scientifica 1},
             city={Roma},
             postcode={00133},
             state={Italy},
             country={}}

\affiliation[22]{organization={Universidad Iberoamericana, Dept. de Fisica y Matematicas},
             addressline={},
             city={Mexico City},
             postcode={01210},
             state={Mexico},
             country={}}

\affiliation[23]{organization={Korea University},
             addressline={145 Anam-ro},
             city={Seongbuk-gu, Seoul},
             postcode={},
             state={Korea},
             country={}}

\affiliation[24]{organization={Université Paris-Saclay},
             addressline={3 rue Joliot Curie, Bâtiment Breguet},
             city={Gif-sur-Yvette},
             postcode={91190},
             state={France},
             country={}}

\begin{abstract}

The RPCs employed at the LHC experiments are currently operated in avalanche mode, with a mixture containing a large fraction of C$_{2}$H$_{2}$F$_{4}$ ($\approx$90\% or more) with the addition of i-C$_{4}$H$_{10}$ and SF$_{6}$ in different concentrations. However, C$_{2}$H$_{2}$F$_{4}$ and SF$_{6}$ are fluorinated greenhouse gases (F-gases) with Global Warming Potential (GWP) of $\approx$1400 and $\approx$22800, respectively. EU regulations imposed a progressive phase-down of C$_{2}$H$_{2}$F$_{4}$ production and consumption, aiming at strongly reducing its emission. This is already resulting in an increase of its price and reduction in availability.

The most desirable long-term solution to this problem is to find an alternative, F-gases-free gas mixture, able to maintain similar detector performance. To address this challenge, the RPC ECOGasas@GIF++ collaboration (including RPC experts of ALICE, ATLAS, CMS, SHiP/LHCb, and the CERN EP-DT group) was created in 2019. The collaboration is currently studying a gas from the olefine family, the C$_{3}$H$_{2}$F$_{4}$ (or simply HFO, with GWP $\approx$6), to be used, in combination with CO$_{2}$, as a substitute for C$_{2}$H$_{2}$F$_{4}$.

This contribution will focus on the signal shape studies that have been carried out by the collaboration during dedicated beam test periods. The methodology used in the data analysis will be presented, together with the results obtained with several HFO-based gas mixtures, and with the currently employed one. Furthermore, results on the counting-rate dependence of the RPC performance, obtained by combining the muon beam with the GIF++ $^{137}$Cs source with different attenuation factors, will also be presented.
\end{abstract}


\end{frontmatter}

\section{Introduction}
\label{sec:intro}

Resistive Plate Chambers (RPCs), are gaseous detectors with planar geometry and resistive electrodes (made of either High Pressure Laminates (HPL) or glass). Thanks to their relatively low-cost, $\approx$ns time resolution and $\approx$mm spatial resolution, they are currently employed in the muon trigger/identification systems of the majority of the LHC experiments \cite{muonTDR,cmsTDR,atlasTDR} and are being considered as a possibility in the LHCb phase II upgrade \cite{lhcBphaseII}. Proper operation of these detectors is granted by the usage of optimised gas mixtures, containing $\approx$90\% C$_{2}$H$_{2}$F$_{4}$ and $<$~1\% SF$_{6}$ (plus a fraction of i-C$_{4}$H$_{10}$ as photon quencher). Although these mixtures satisfy all the performance requirements, they contain a high fraction of C$_{2}$H$_{2}$F$_{4}$ and SF$_{6}$, which are classified as fluorinated greenhouse gases (F-gases/GHGs).

With the 2017 EU F-gases regulations \cite{euReg}, enforced once more in 2024 \cite{euReg2}, a progressive phase-down in production and usage of these gases has been imposed, leading to an increase in cost and reduction in availability. For this reason, CERN has adopted a policy of F-gases usage reduction and, since the currently employed RPC gas mixture is almost entirely made up of high-GWP F-gases and it represent a significant fraction of the total GHG emission of the LHC experiments\cite{beatrice}, it is of utmost importance to search for a more eco-friendly RPC gas mixture. Following several independent laboratory studies \cite{prelGiorgia,prelAntonio,prelGianluca,prelPiccolo}, the RPC ECOGas@GIF++ collaboration (including researchers from ALICE, ATLAS, CMS, LHCb/SHiP and the CERN EP-DT group) was created to join forces among RPC experts of the different LHC experiments, sharing knowledge and person-power. The main goal and the results obtained so far by the collaboration have been discussed extensively in \cite{ecogas,ecogasThin,focusPoint}. The studies focus on fully replacing the C$_{2}$H$_{2}$F$_{4}$ with one of its industrial substitute, the \textit{tetrafluoropropene} (C$_{3}$H$_{2}$F$_{4}$ or simply HFO), in its -ze isomer, diluted with CO$_{2}$, to lower the detector working voltage. 

It has been pointed out that HFO could potentially dissociate in the high atmosphere, leading to the creation of trifluoroacetic acid (TFA, a compound harmful to humans in high concentrations) which could precipitate thanks to rainfall \cite{pfas1,pfas2}. A debate has been ongoing on the matter and the current outcome is that the actual impact should be irrelevant (as described in \cite{pfas3}). This is, nevertheless, a potential issue to be considered and it will probably require deeper investigation in the future. 

This paper reports the main results obtained from the signal shape studies and RPC response evolution when exposed to different background irradiation levels induced by a $^{137}$Cs source and operated with several HFO/CO$_{2}$-based gas mixtures, in the context of the RPC ECOGas@GIF++ collaboration. The document is divided as follows: Section \ref{sec:setup} contains a description of the experimental setup and the methodology used in the data-taking/analysis, Section \ref{sec:results} reports the main results obtained in terms of signal shapes and RPC response evolution for increasing background rate (in \ref{sub:signal}) as well as performance evolution throughout the ongoing long-term operation studies (in \ref{sub:perfEvol}). Lastly, Section \ref{sec:conclusion} is dedicated to the conclusions and to possible outlooks for the future of this work.

\section{Experimental setup}
\label{sec:setup}

The experimental setup of the collaboration is located at the CERN Gamma Irradiation Facility (GIF++)\cite{GIF++}, located on the H4 secondary SPS beam line. This facility is equipped with a high activity $^{137}$Cs source ($\approx$12.5~TBq) and, during dedicated beam time periods, it is traversed by a high energy muon beam. The latter can be used to study the detector performance while the former allows one to induce a high radiation background on the detectors under test. This allows one to simulate long-term operation periods in a much shorter time-span (\textit{aging studies}) and, when combined with the muon beam, to study the detectors response under increasing background levels (rate capability). The irradiation from the $^{137}$Cs source can be modulated by means of lead attenuation filters and a total of 27 possible irradiation intensities can be obtained.

Table \ref{tab:detectors} reports the main features of all the detectors of the collaboration. Note that the CMS RE11 and KODEL-H RPCs have a double gas gap. Table \ref{tab:detectors} also includes a column related to the readout used for each RPC; this is important since this contribution will only focus on results from the ALICE and EP-DT detectors because they are equipped with a digitizer for raw signal readout. For a more detailed description of the results obtained with the other detectors, the reader can refer to \cite{AbbrProc}.

\begin{table}[h!]
\footnotesize
	\begin{center}
		\setlength{\tabcolsep}{1.3pt} 
        \begin{tabular}{cccccc}\\\toprule  
        \textbf{Name} & \textbf{Gaps} & \textbf{Gap [mm]} & \textbf{Electrode [mm]} & \textbf{Area [cm$^{2}$]} & \textbf{Readout]} \\\midrule
        ALICE & 1 & 2 &  2 & 2500 & Digitizer \\  \midrule
        ATLAS & 1 & 2 & 2 & 550 & Digitizer  \\  \midrule
        \multirow{2}{5em}{CMS RE11} &  \multirow{2}{4em}{\centering 2} & \multirow{2}{4em}{\centering 2} & \multirow{2}{4em}{\centering 2} & 3627( & TDC  \\ 
         & & & & 4215 & TDC \\ \midrule
        EP-DT & 1 & 2 & 2 & 7000 & Digitizer \\  \midrule
        LHCb/SHiP & 1 & 1.6 & 1.6 & 7000 & TDC \\ \midrule
        \multirow{2}{5em}{KODEL-H} &  \multirow{2}{4em}{\centering 2} &  \multirow{2}{4em}{\centering 1.4} &  \multirow{2}{4em}{\centering 1.4} & 2500 & TDC \\ 
        & & & & 2500 & TDC  \\ \bottomrule
        \end{tabular}
        \caption{Features of the RPC ECOGas@GIF++ collaboration detectors}\label{tab:detectors}
	\end{center}
\end{table}

\begin{figure}[h!]
\includegraphics[width=\linewidth]{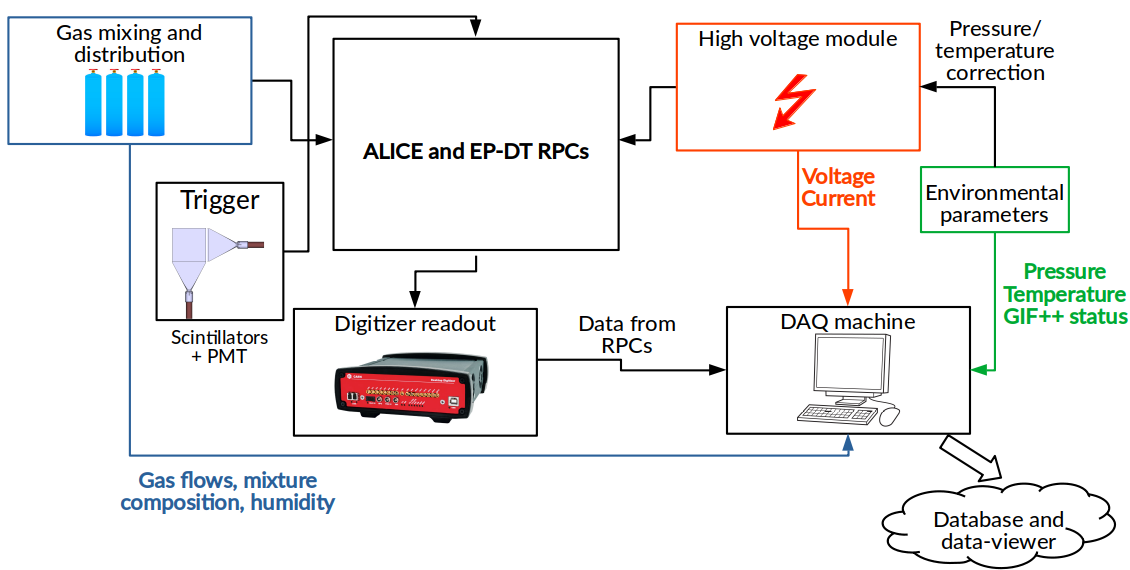}
\caption{Sketch of the experimental setup installed in GIF++. A detailed description of each component is reported in the text}
\label{fig:setup}
\end{figure}

Figure \ref{fig:setup} shows a sketch of the experimental setup currently installed at the GIF++. The gas mixing and distribution system allows one to mix up to four gases in the desired concentrations. Moreover, it also allows one to regulate the relative humidity of the mixture by changing the amount of gas flowing in a humidifier tank. The relative humidity of the mixture is set to 40\% to keep the bakelite resistivity at a constant value. The high voltage (HV) to power the detectors is provided by means of a CAEN SY1527 mainframe\footnote{\url{https://www.caen.it/subfamilies/mainframes/}} and two CAEN A1526 boards\footnote{\url{https://www.caen.it/products/a1526/}} (one with positive and one with negative polarity). A monitoring software is employed to continuously store all the relevant parameters (i.e. environmental conditions, gas mixture composition, current absorbed by the detectors and applied high voltage) on a dedicated database for later analysis. During the beam tests, a scintillator-based trigger is also installed and the data from the detectors are acquired by the readout system.

\subsection{Methodology and analysis}
\label{sub:methodology}

During the beam tests, the trigger for the beam, is provided by a four-fold coincidence of scintillators (two installed by the RPC ECOGas@GIF++ collaboration inside the irradiation bunker and two installed by the GIF++ team outside) that results in an effective trigger area of $\approx$10$\times$10~cm$^{2}$. 

In a usual \textit{HV scan}, 12 increasing HV values are scanned and, for each voltage applied to the detectors, at least 2500 beam-induced triggers are collected, to measure the RPC efficiency. The RPCs can also be used to provide an estimate of the radiation background rate; indeed this measurement is biased by the low $\gamma$ detection efficiency of the RPCs (in the order of the \textperthousand) but it is nonetheless useful to compare the results obtained with the different mixtures and with different attenuation factors. Two different techniques are adopted for the ALICE and EP-DT detectors and they will be described in the following. 

For each mixture, a fast ($\approx$200~triggers per HV value) scan is performed without any background (\textit{source-off}) to fine tune the values of the voltages to be scanned. Following this, a longer source-off scan is taken. Typically, 4/5 scans with increasing background levels (\textit{source-on}) are carried out by changing the filters used to shield the $\gamma$ source. A quick data-analysis is carried out before moving on, to possibly add other HV values.   

Several mixtures have been beam-tested by the RPC ECOGas@GIF++ collaboration and their composition is reported in Table \ref{tab:mixtures}, together with the gas mixture currently employed by ATLAS and CMS \cite{stdCMS, stdATLAS} (STD), taken as a reference to which the eco-friendly alternatives will be compared.

\begin{table}[h!]
\small
	\begin{center}
		\setlength{\tabcolsep}{1.25pt} 
        \begin{tabular}{ccccccc}\\\toprule  
        \textbf{Name} & \textbf{C$_{2}$H$_{2}$F$_{4}$} & \textbf{HFO} & \textbf{CO$_{2}$} & \textbf{i-C$_{4}$H$_{10}$} & \textbf{SF$_{6}$} & \textbf{GWP} \\\midrule
        STD & 95.2 & 0 &  0 & 4.5 & 0.3 & 1488 \\  \midrule
        MIX0 & 0 & 0 & 95 & 4 & 1 & 730\\  \midrule
        MIX1 & 0 & 10 & 85 & 4 & 1 & 640\\  \midrule
        MIX2 & 0 & 20 & 75 & 4 & 1 & 560\\  \midrule
        MIX3 (ECO3) & 0 & 25 & 69 & 5 & 1 & 529\\  \midrule
        MIX4 & 0 & 30 & 65 & 4 & 1 & 503\\  \midrule
        MIX5 (ECO2) & 0 & 35 & 60 & 4 & 1 & 482\\  \bottomrule
        MIX6 & 0 & 40 & 55 & 4 & 1 & 457 \\  \bottomrule
        \end{tabular}
        \caption{Percentage-wise composition of the gas mixtures tested}
        \label{tab:mixtures}
	\end{center}
\end{table}

\begin{figure}[h!]
\centering
\includegraphics[width=0.85\linewidth]{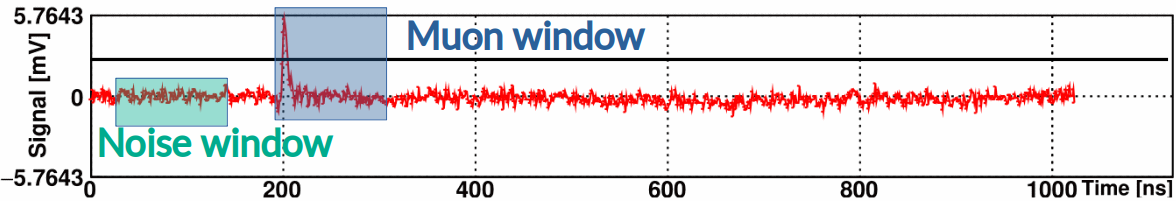}
\caption{Example of a signal obtained with the ALICE detector, using the standard gas mixture. The muon and noise windows (described in the text) are highlighted in the figure.}
\label{fig:exSignal}
\end{figure}

Figure \ref{fig:exSignal} shows an example of a waveform obtained with the ALICE detector, when operated with the STD gas mixture. The signal is sampled at 1~Gs/s and the number of samples is fixed to 1024 by the design of the utilized digitizer (\href{https://www.caen.it/products/dt5742/}{CAEN DT5742} with 1~V peak-to-peak input signal amplitude, 12-bit vertical resolution and sampling frequency of 1/2.5/5~Gs/s). For the data processing, two distinct time intervals are highlighted: the \textit{muon window}, in the range 190-250~ns and the \textit{noise window}, in the range 70-150~ns. The former is a time interval in which the signal due to the presence of the muon is expected while in the latter no signal (except for the noise) should be present. The threshold used to consider the chamber as efficient for a given trigger has been set to 5 times the RMS of the signal in the noise window (marked with an horizontal black line in Figure \ref{fig:exSignal}).

In the case of the EP-DT detector, the threshold has been fixed to 2~mV, independently of the mixture and the scan (for the ALICE detector, the average threshold obtained with the procedure described above is $\approx$1.6~mV). Also, a different digitizer (\href{https://www.caen.it/products/v1730/}{CAEN V1730} with 2~V peak-to-peak input signal amplitude, 14-bit vertical resolution and sampling frequency of 0.5~Gs/s) was utilized but similar definitions for the muon and noise windows have been adopted. For both RPCs, 7 strips were read out, horizontal for EP-DT and vertical for ALICE, with a pitch of $\approx$2.1~cm.

The rest of the analysis, performed independently by the ALICE and EP-DT groups, is aimed at characterizing the detector response. In particular, for each trigger, the algorithm runs over the data from all the strips and checks whether the amplitude of the signal in the muon window is above the threshold. Following this step, the algorithm also checks whether the signal is actually induced by a muon or by a possible cross-talk between adjacent strips. In particular, it has been noticed that in this case, the polarity of the peak in the muon window is opposite with respect to the expected one and/or is characterized by two consequent peaks with opposite polarities. This category of events is then discarded from further processing.

If a signal is originating from a muon hit it is processed further and the analysis algorithm computes other parameters such as: the signal prompt charge as the integral in time of the fast electron signal, the time-over-threshold (the time interval during which the signal amplitude is above threshold), and the cluster size (the number of adjacent strips that have been fired in any given trigger). For the signals induced by gammas, the cluster rate is also computed as the ratio between the number of detected photons and the product between the run duration and the area of the detector covered by the strips, and it is measured in $Hz/cm^{2}$. In the case of the EP-DT detector, the rate of impinging gammas is computed by opening a long ($\approx$1.2~ms) acquisition window, during which the number of peaks measured (in absence of a beam-induced trigger) is used as an estimate of the number of $\gamma$ hits and is then divided by the length of the acquisition window and the area, to have an estimate of the $\gamma$ rate.

Once the analysis has been run on a full scan, it also produces the trend of the aforementioned vales as a function of the effective high voltage (HV$_{\text{eff}}$, which takes into account the variation of detector gain depending on temperature and pressure) \cite{ecogas} to fully characterize the detector response with a given mixture and for a given irradiation condition.

\section{Results}
\label{sec:results}

This section reports a selection of the main results of the signal shape studies carried out during beam test periods at the GIF++, as well as the evolution of the RPC performance throughout the ongoing aging campaign.

\subsection{Signal shape studies}
\label{sub:signal}

\begin{figure}[h]
    \centering
    \begin{subfigure}[t]{0.5\linewidth}
        \centering
        \includegraphics[width = \linewidth]{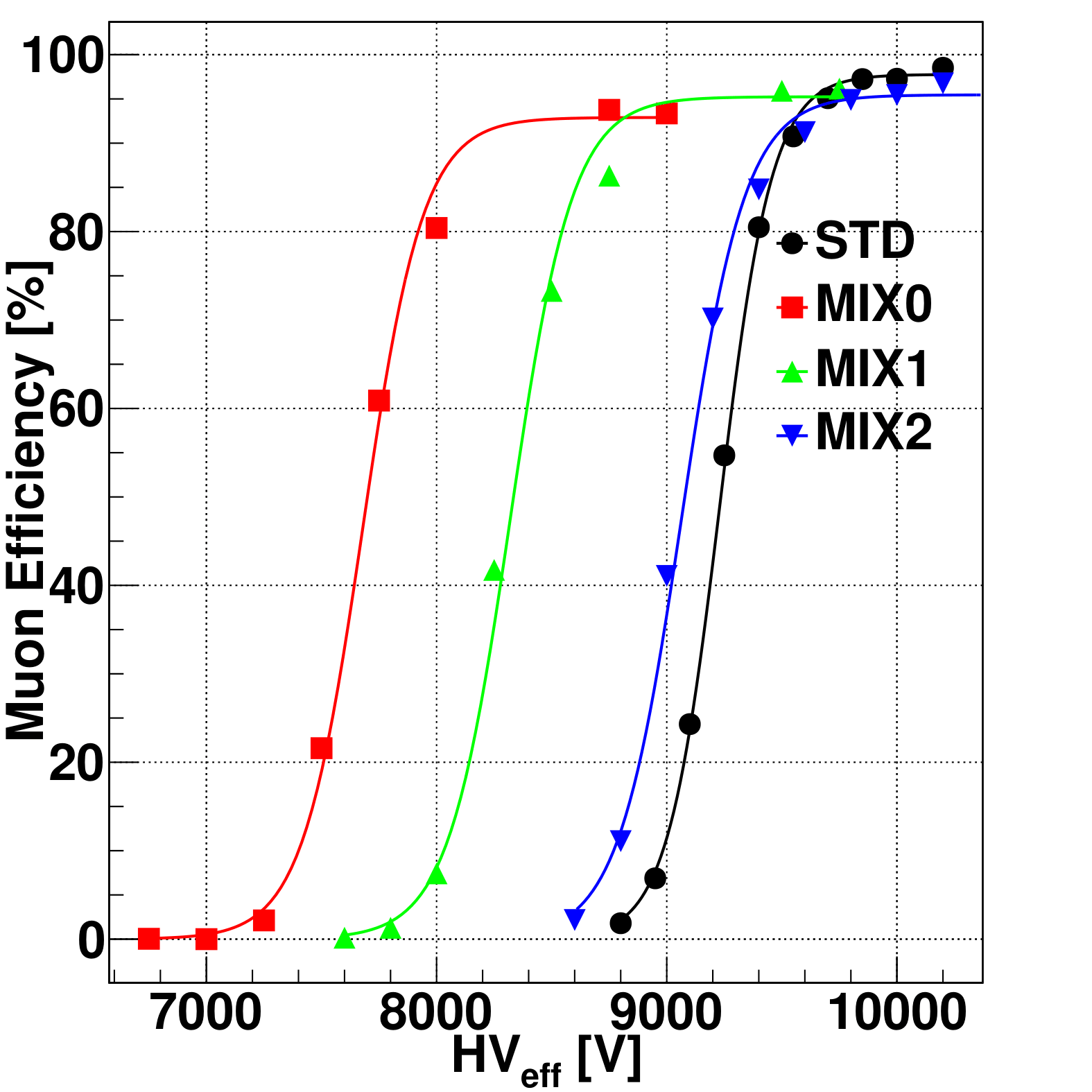}
        \caption{ALICE detector}
        \label{fig:aliceEffSourceOff}
    \end{subfigure}%
    \hfill
    \begin{subfigure}[t]{0.5\linewidth}
        \centering
        \includegraphics[width = \linewidth]{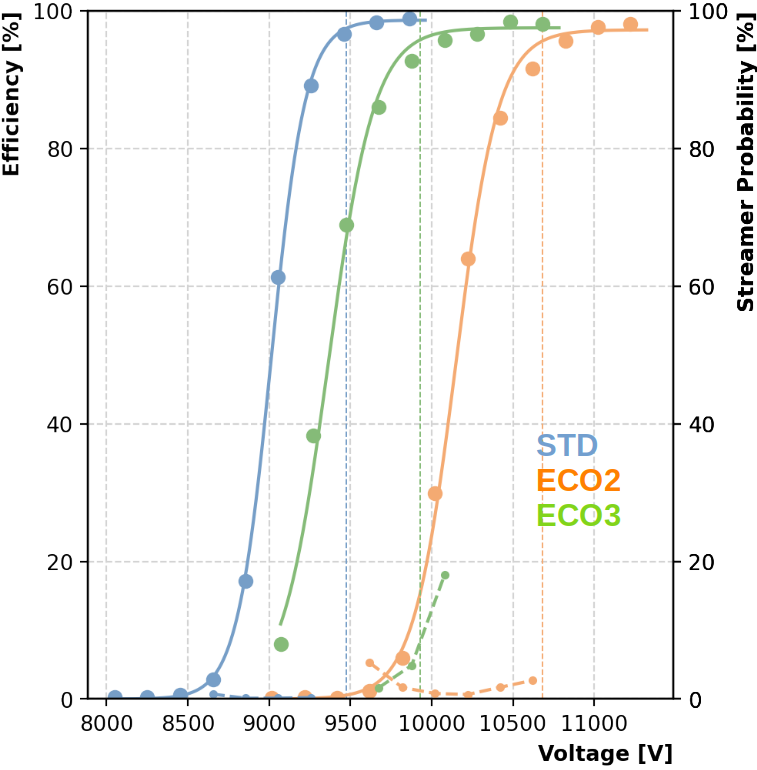}
        \caption{EP-DT detector}
        \label{fig:epdtEffSourceOff}
    \end{subfigure}
    \caption{Source-off efficiency vs HV$_{\text{eff}}$ curves for the ALICE (left panel) and EP-DT (right panel) RPCs}
    \label{fig:effSourceOff}
\end{figure}

The first result obtained is the trend of the muon detection efficiency as a function of the HV$_{\text{eff}}$ at source-off (i.e. without any background irradiation). Figures \ref{fig:aliceEffSourceOff} and \ref{fig:epdtEffSourceOff} show the source-off efficiency for the ALICE and EP-DT detectors respectively, when flushed with the different gas mixtures, including STD, shown with black (blue) markers for ALICE (EP-DT).

The efficiency curves have been interpolated using a logistic function (as usually done for RPCs, as reported for example in \cite{ecogas}), to extract information about the detector working point (WP, i.e. the voltage at which full efficiency is reached), defined as the knee voltage (HV$_{\text{eff}}$ where the efficiency reaches 95\% of its maximum) plus 150~V. The results show that increasing the HFO concentration in the mixture leads to an increase of the WP by $\approx$1~kV for every 10\% HFO added. Moreover, a slight increase of the maximum efficiency is observed when more HFO is added to the gas mixture. 

\begin{figure}[h]
    \centering
    \begin{subfigure}[t]{0.5\linewidth}
        \centering
        \includegraphics[width = \linewidth]{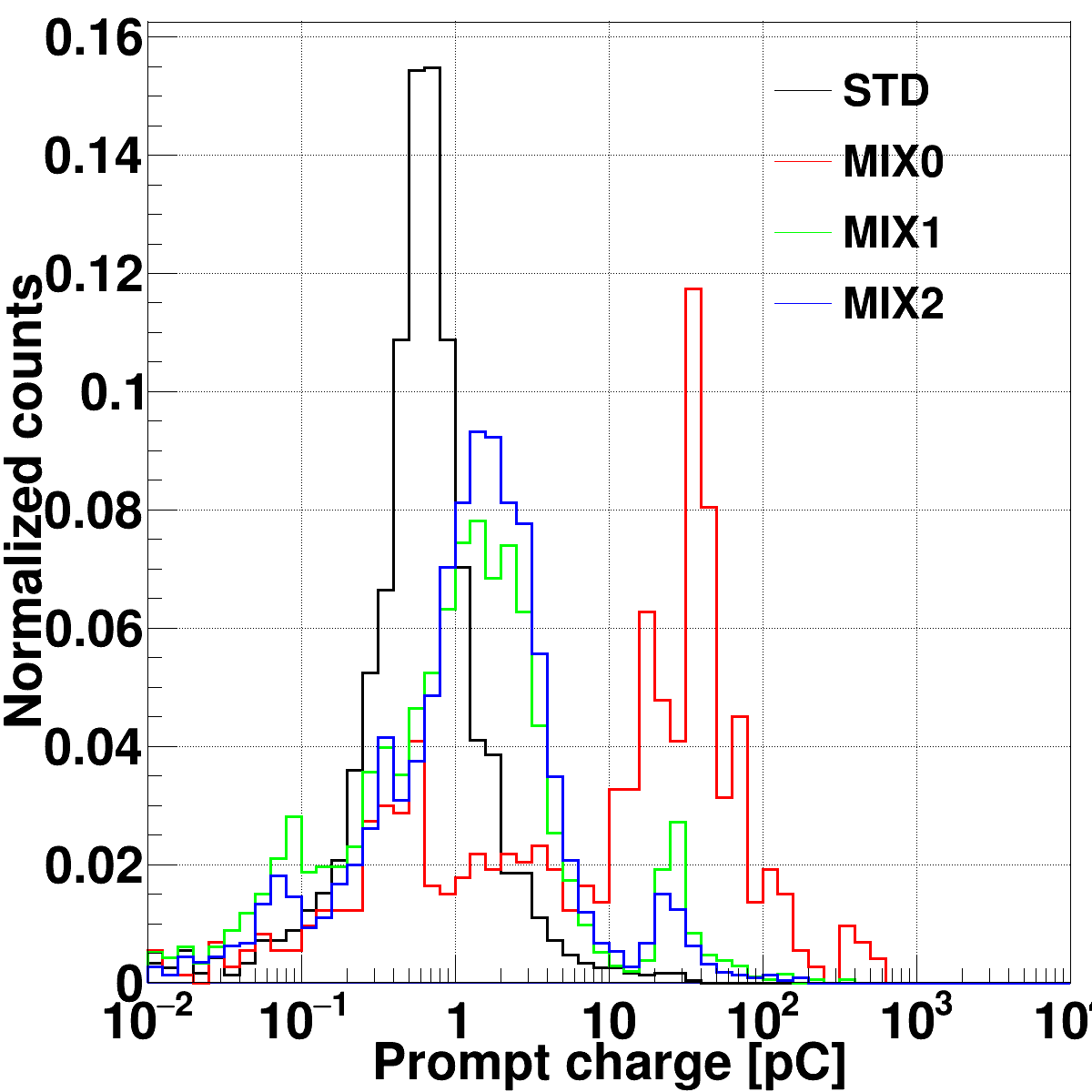}
        \caption{ALICE detector}
        \label{fig:chargeALICEoff}
    \end{subfigure}%
    \hfill
    \begin{subfigure}[t]{0.5\linewidth}
        \centering
        \includegraphics[width = 0.94\linewidth]{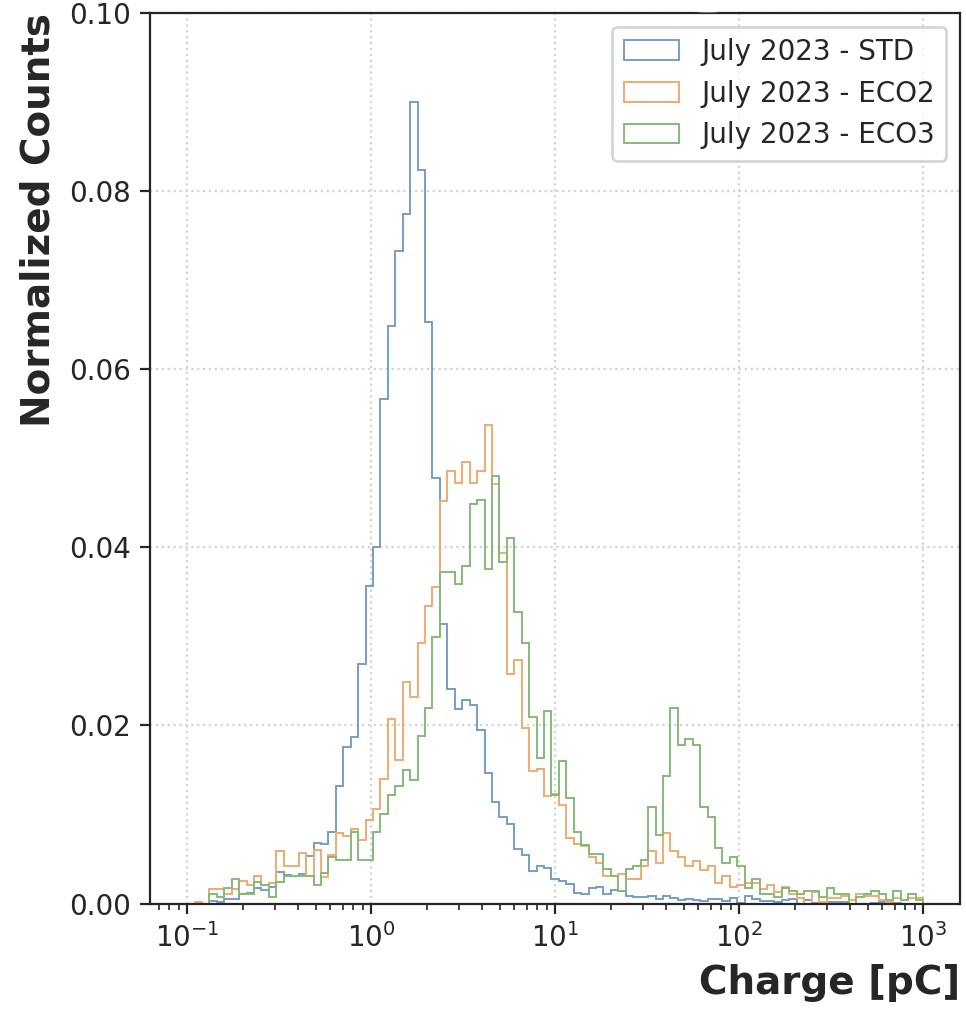}
        \caption{EP-DT detector}
        \label{fig:chargeEpdtOff}
    \end{subfigure}
    \caption{Left panel: Source-off prompt charge distributions for the ALICE RPC. Right panel: EP-DT RPC}
    \label{fig:ChargeOff}
\end{figure}

Figures \ref{fig:chargeALICEoff} and \ref{fig:chargeEpdtOff} show the source-off prompt charge distribution obtained at the voltage closest to the estimated working point for some of the tested gas mixtures, including STD, shown with black (blue) lines for ALICE (EP-DT).

One can clearly see that, for all the eco-friendly alternatives, the avalanche peak (the peak at lower values of the spectrum, populated by the pure \textit{avalanche signals}) is shifted to higher values, with respect to the STD gas mixture. A similar observation can be drawn for what concerns the number of large signals (peak at higher values of the spectrum, populated by the signals that lead to a larger charge release in the gas, for example streamers), which is more populated than the one of the STD gas mixture. Increasing the HFO fraction in the mixture seems to mitigate both effects, both by reducing the number of large signals as well as shifting the average avalanche peak to lower values. All in all, increasing the HFO concentration in the mixture seems to improve the RPC response. It is worth pointing out that, as shown in \cite{focusPoint}, the large-signals contamination increases with HV$_{\text{eff}}$ more rapidly than in the case of the STD mixture, making the \textit{good} operating region narrower for these mixtures. 

\begin{figure}[h]
    \centering
    \begin{subfigure}[t]{0.5\linewidth}
        \centering
        \includegraphics[width = \linewidth]{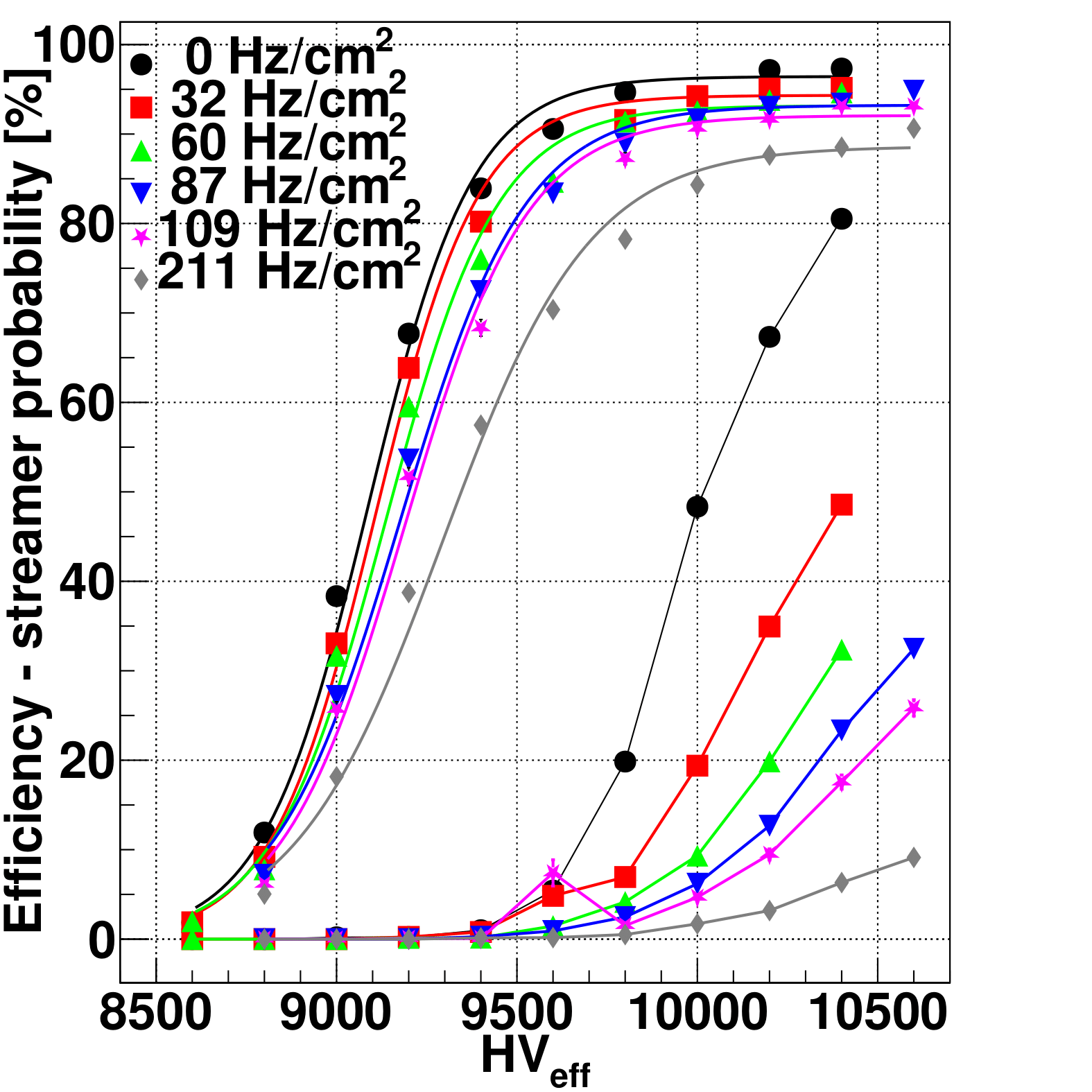}
        \caption{ALICE detector}
        \label{fig:aliceEffSourceOn}
    \end{subfigure}%
    \hfill
    \begin{subfigure}[t]{0.5\linewidth}
        \centering
        \includegraphics[width = 0.93\linewidth]{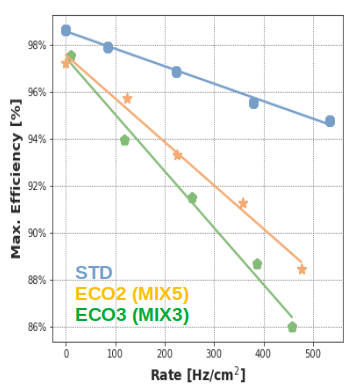}
        \caption{EP-DT detector}
        \label{fig:epdtEffSourceOn}
    \end{subfigure}
    \caption{Left panel: Efficiency/large signal probability vs HV$_{\text{eff}}$ curve for the ALICE RPC when flushed with MIX2 (20/75 HFO/CO$_{2}$) for different background levels. Right panel: Maximum efficiency for the EP-DT RPC as a function of the measured $\gamma$ cluster rate}
    \label{fig:sourceONEx}
\end{figure}

When the RPCs are exposed to the $\gamma$ source, one can study how the response to the muon beam evolves for increasing irradiation levels. Figure \ref{fig:aliceEffSourceOn} shows the efficiency and large signal contamination as a function of HV$_{\text{eff}}$ for the ALICE RPC, when flushed with one of the tested eco-friendly gas mixtures (MIX2, see Table \ref{tab:mixtures}) for different irradiation levels (quantified by the $\gamma$ cluster rate measured by the RPC itself). Three main effects are visible: the efficiency curves are shifted to higher voltages with respect to the source-off condition, the maximum value of efficiency decreases with increasing irradiation and there is a reduction of the large signal contamination. These effects can be explained, at least partially, by considering that, when the irradiation increases, so does the absorbed current. This current is flowing through all the elements of the RPC, including the bakelite electrodes which behave as resistors and across which the current leads to a sizeable voltage drop, effectively leading to a decrease of the voltage applied to the gas and a consequent reduction of the RPC gain. As explained in \cite{ecogas}, this effect can be accounted for by subtracting from HV$_{\text{eff}}$ the voltage drop on the bakelite electrodes, estimated as the product of the circulating current and the electrode resistance. Figure \ref{fig:epdtEffSourceOn} shows the maximum efficiency for the EP-DT detector, for three mixtures and several irradiation levels (notice that the maximum rate measured by this RPC is higher than the one measured by ALICE since it is located closer to the irradiation source).

\begin{figure}[h]
    \centering
    \begin{subfigure}[t]{0.5\linewidth}
        \centering
        \includegraphics[width = \linewidth]{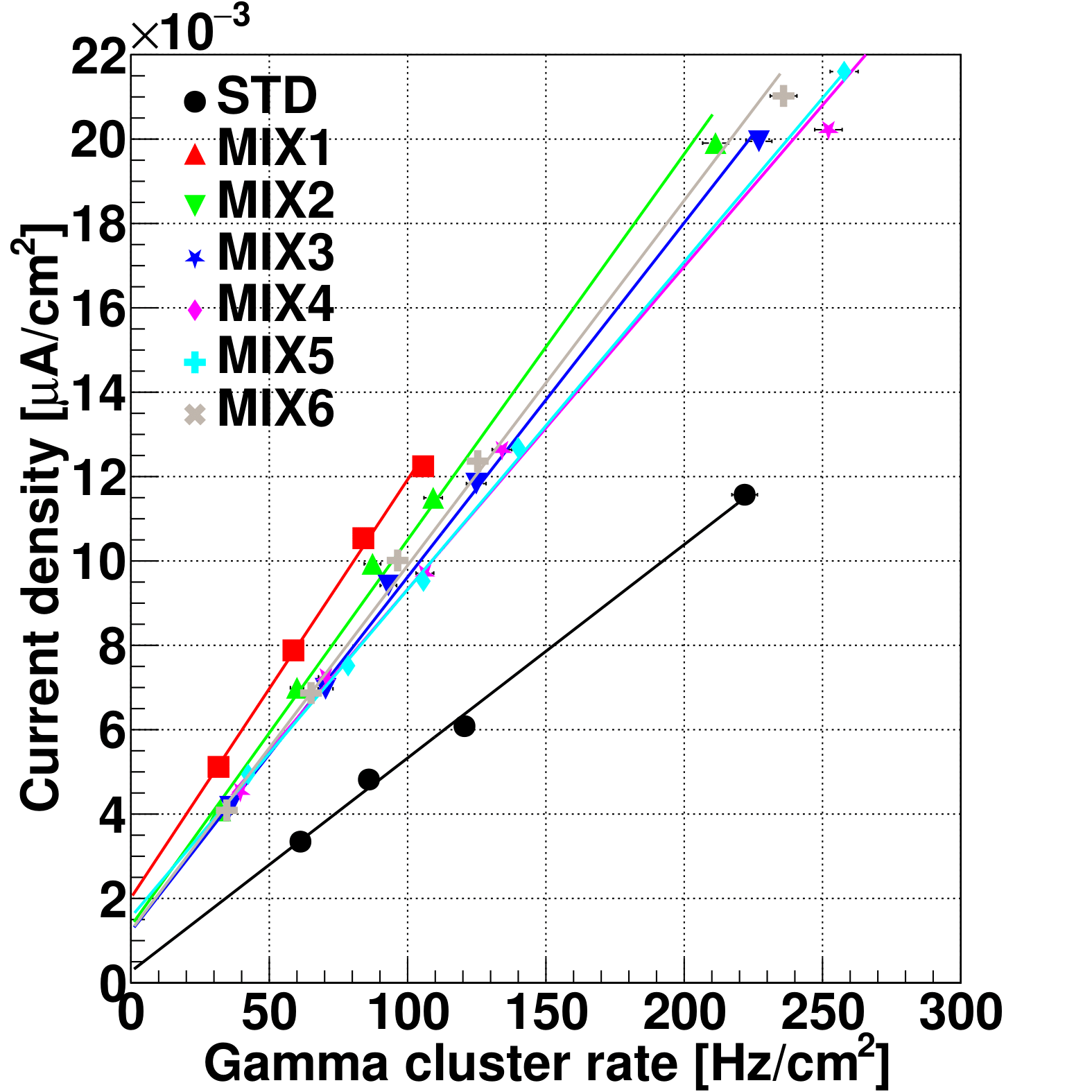}
        \caption{Current vs $\gamma$ cluster rate}
        \label{fig:sourceOnCurr}
    \end{subfigure}%
    \hfill
    \begin{subfigure}[t]{0.5\linewidth}
        \centering
        \includegraphics[width = \linewidth]{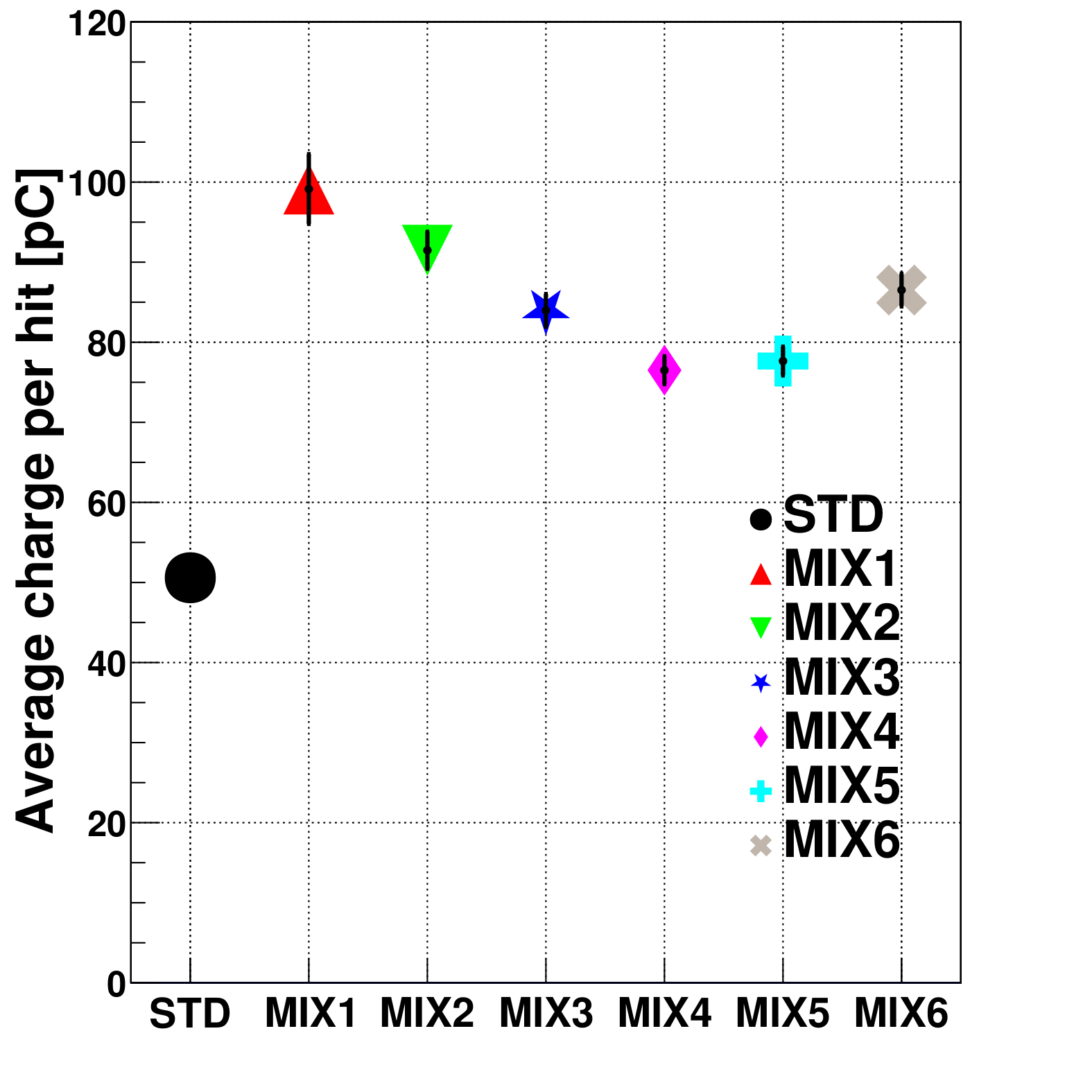}
        \caption{Average charge per $\gamma$ cluster}
        \label{fig:chargePerHit}
    \end{subfigure}
    \caption{Left panel: Current density vs measured $\gamma$ cluster rate for the ALICE RPC for all the tested gas mixtures. The straight lines shown in the panels are linear fits of the data. Right panel: Estimated average charge released in the gas per $\gamma$ hit for all the tested gas mixtures}
    \label{fig:currAndCharge}
\end{figure}

The relationship between the circulating current and the detected $\gamma$ rate can be expressed as:

    \begin{equation}
        \label{eq:currRate}
        \frac{I}{A} =  \langle Q \rangle \cdot \frac{N_{\gamma-detected}}{A \cdot \Delta t} + DCD
    \end{equation}

where $\frac{I}{A}$ is the current per unit area, $\langle Q \rangle$ is the average charge per hit, $\frac{N_{\gamma-detected}}{A \cdot \Delta t}$ is the measured $\gamma$ cluster rate and $DCD$ is the \textit{dark current} per unit area (i.e. the current flowing when the detector is not exposed to any particle source). This relationship can be exploited as shown in Figure \ref{fig:currAndCharge}, where the left panel shows the trend of the absorbed current density (at the WP estimated for the specific irradiation condition) as a function of the measured $\gamma$ cluster rate for all the tested mixtures and the parameter $\langle Q \rangle$ is extracted with a linear fit. Its value, computed for all the gas mixtures under test, is reported in the right panel of Figure \ref{fig:currAndCharge}. This is an estimation of the average total charge released in the gas and, by looking at Figure \ref{fig:chargePerHit} one can infer that, on average, this quantity is $\approx$1.5/2 times higher for the eco-friendly alternatives with respect to the STD gas mixture. This higher charge per hit could, in principle, lead to a faster detector aging and this hypothesis is being closely investigated with a dedicated long-term aging campaign.  

\subsection{Performance evolution during aging studies}
\label{sub:perfEvol}

As anticipated earlier, an aging study is ongoing to assess the long-term evolution of the RPC response when operated with HFO-based gas mixtures. Among all the tested candidates, the one referred to as "ECO2" (or MIX5) has been selected, thanks to its good trade-off between detector performance and increase in working voltage. For a more in-depth discussion on the aging test methodology and main results, the reader can refer to \cite{AbbrProc}, while only some results obtained with the EP-DT detector will be discussed here.

\begin{figure}[h]
    \centering
    \begin{subfigure}[t]{0.5\linewidth}
        \centering
        \includegraphics[width = \linewidth]{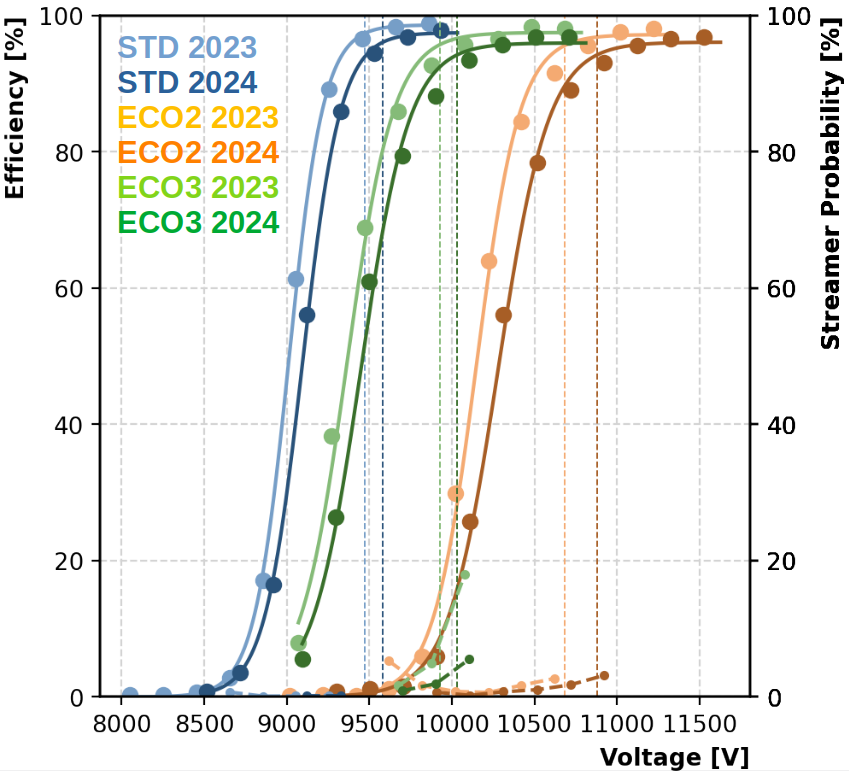}
        \caption{EP-DT RPC, source-off comparison}
        \label{fig:epdtSourceOffEvolution}
    \end{subfigure}%
    \hfill
    \begin{subfigure}[t]{0.5\linewidth}
        \centering
        \includegraphics[width = 0.95\linewidth]{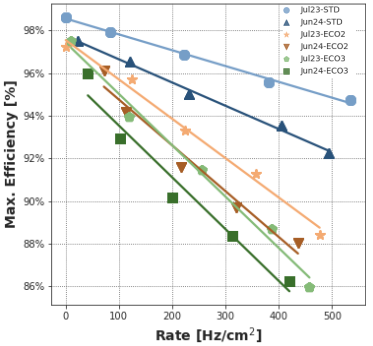}
        \caption{EP-DT RPC, source-on comparison}
        \label{fig:epdtSourceOnEvolution}
    \end{subfigure}
    \caption{Left panel: Source-off efficiency and large signal contamination curves for the EDP-DT RPC and for three different gas mixtures. Comparison of the results obtained in July 2023 (before the start of the aging campaign) and July 2024, after the integration of $\approx$115~mC/cm$^{2}$. Right panel: Comparison of the maximum efficiency as a function of the measured $\gamma$ cluster rate in the same time frame}
    \label{fig:epdtAging1}
\end{figure}

Figure \ref{fig:epdtSourceOffEvolution} shows the comparison between the efficiency and large signal probability vs HV$_{\text{eff}}$ curves at source-off before the start of the aging studies for this specific detector (July 2023) and after one-year exposure to the $^{137}$Cs source (July 2024), corresponding to an integrated charge of $\approx$115~mC/cm$^{2}$. Figure \ref{fig:epdtSourceOnEvolution} shows instead the comparison of the maximum efficiency as a function of the $\gamma$ rate in the same time frame.

The source-off comparison shows an increase of the detector working point $\approx$100~V for the STD gas mixture and $\approx$150/200~V for ECO3 and ECO2 respectively. This is also accompanied by a decrease of the maximum efficiency of $\approx$2\% for all the mixtures under test and this can be attributed to a a geometrical misalignment of the detector in 2024, not present in 2023. Lastly, the large signal probability seems to be reduced for all the three mixtures. Concerning the source-on results, the maximum efficiency under irradiation seems to be reduced by 2\% for all mixtures and for all the irradiation conditions. 

\begin{figure}[h]
    \centering
    \begin{subfigure}[t]{0.5\linewidth}
        \centering
        \includegraphics[width = 0.95\linewidth]{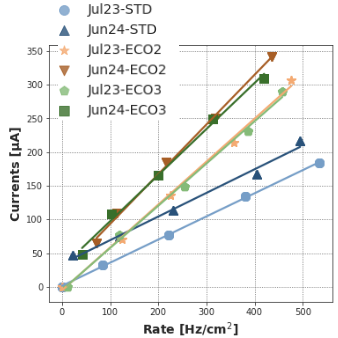}
        \caption{EP-DT RPC, currents at source-on}
        \label{fig:epdtSourceOnCurrComparison}
    \end{subfigure}%
    \hfill
    \begin{subfigure}[t]{0.5\linewidth}
        \centering
        \includegraphics[width = \linewidth]{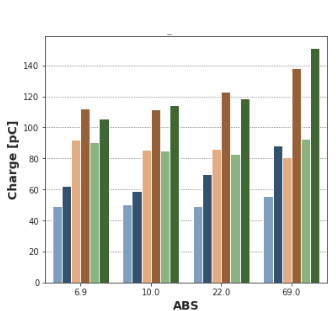}
        \caption{EP-DT RPC, average charge per $\gamma$ hit}
        \label{fig:epdtSourceOnChargeComparison}
    \end{subfigure}
    \caption{Left panel: Comparison of the absorbed current at working point as a function of the measured $\gamma$ rate between July 2023 and July 2024, after the integration of $\approx$115~mC/cm$^{2}$. The straight lines represent a linear fit to the data. Right panel: Comparison of the average charge per $\gamma$ hit for all the mixtures and all the irradiation conditions in July 2023 and 2024}
    \label{fig:epdtAging2}
\end{figure}

Figure \ref{fig:epdtSourceOnCurrComparison} shows the comparison of the current measured at the detector working point as a function of the measured background rate between July 2023 and July 2024, for the EP-DT detector. The currents are slightly higher in 2024 and this can be attributed, at least partly, to an observed increase of the detector dark current. This could be potentially related to a degradation of the electrodes surface but this can only be proven with dedicated chemical analyses which have not yet been performed. The ratio between the measured current and $\gamma$ rate gives an estimate of the average charge per hit. These results and the comparison between July 2023 and 2024, are shown in Figure \ref{fig:epdtSourceOnChargeComparison}. The values reported on the x-axis correspond to the value of the filter (absorption factor/ABS) used to obtain a specific background rate. The latter goes from 1 (maximum irradiation) to 46000 (minimum irradiation). In general, the average charge per hit is higher for all the mixtures and for all the ABS values in 2024 with respect to 2023 and, once more, this can be partly explained by considering the aforementioned dark current increase.

\section{Conclusions and outlook}
\label{sec:conclusion}

The search for an eco-friendly alternative gas mixture for RPC detectors is a hot-topic in the RPC detector community, especially due to the EU F-gases regulations, which foresee a progressive phase down in the production and usage of F-gases.

Several eco-friendly alternatives, where C$_{2}$H$_{2}$F$_{4}$ is replaced by different concentrations of HFO/CO$_{2}$ have been identified using cosmic-ray tests and they are currently being characterized further, by means of beam tests and rate capability studies, by the RPC ECOGas@GIF++ collaboration.

The main takeaway message of the studies carried out so far, is that the RPC response improves if the HFO concentration in the mixture is increased. On the flip side, the working point of the detectors also increases and, for this reason, a proper balance must be found. In general, the average charge released at working point in the gas per hit is $\approx$1.6/1.7 times higher with respect to the one of the currently employed gas mixture; the possible aging effects due to the higher charge are being carefully studied by the RPC ECOGas@GIF++ collaboration by means of a long-term aging study.

The evolution of the RPC performance is being studied throughout the aging process. The studied ALICE RPC has integrated $\approx$80~mC/cm$^{2}$ while the EP-DT detector has integrated $\approx$115~mC/cm$^{2}$. The former detector showed an increase in absorbed current, muon prompt charge and signal amplitude and these effects are currently being investigated while the latter showed a slight increase of working point and a decrease of the maximum efficiency, without any other significant performance degradation.

The aging campaign, together with further performance evolution studies, is still ongoing and further in-depth analyses of the beam test results are being carried out.


\begin{thebibliography}{99}
    
\bibitem{muonTDR} %
\footnotesize ALICE collaboration, \emph{ALICE Technical Design Report of the Dimuon Forward Spectrometer}, CERN/LHCC 99-22 (\href{https://inspirehep.net/literature/517335}{1999})

\bibitem{cmsTDR} %
\footnotesize CMS collaboration, \emph{The CMS muon project: Technical Design Report}, CERN/LHCC 97-32 (\href{https://cds.cern.ch/record/343814}{1997})

\bibitem{atlasTDR} %
\footnotesize ATLAS collaboration, \emph{ATLAS muon spectrometer: Technical Design Report}, CERN/LHCC 97-022 (\href{https://cds.cern.ch/record/331068}{1997})

\bibitem{lhcBphaseII} 
\footnotesize LHCb collaboration, \emph{Framework TDR for the LHCb Upgrade II: Opportunities in flavour physics, and beyond, in the HL-LHC era}, CERN-LHCC-2021-012 (\href{https://cds.cern.ch/record/2776420}{2021})

\bibitem{euReg} %
\footnotesize Council of European Union, \emph{Council regulation ({EU}) no 517/2014} (\href{https://eur-lex.europa.eu/legal-content/IT/ALL/?uri=CELEX\%3A32014R0517}{2014})

\bibitem{euReg2} %
\footnotesize Council of European Union, \emph{Council regulation ({EU}) no 573/2024} (\href{https://eur-lex.europa.eu/eli/reg/2024/573/oj}{2024})

\bibitem{beatrice}
\footnotesize Mandelli, B. \emph{et al.}, \emph{Strategies for reducing the use of greenhouse gases from particle detectors operation at the CERN LHC Experiments}, Journal of Physics: Conference Series (\href{https://iopscience.iop.org/article/10.1088/1742-6596/2374/1/012159}{2022})

\bibitem{prelGiorgia} %
\footnotesize Proto, G., \emph{Study of the performance of the RPC detector with new eco-friendly gas mixtures}, Il nuovo Cimento C (\href{https://doi.org/10.1393/ncc/i2021-21070-1}{2021})
	
\bibitem{prelAntonio} %
\footnotesize Bianchi, A. \emph{et al.}, \emph{Studies on tetrafluoropropene-based gas mixtures with low environmental impact for Resistive Plate Chambers}, JINST 15 C04039 (\href{https://doi.org/10.1088/1748-0221/15/04/C04039}{2020})

\bibitem{prelGianluca} %
\footnotesize Rigoletti, G. \emph{et al.}, \emph{Characterization of RPC detectors with LHC-like background radiation and new eco-friendly gas mixtures}, JINST 15 C11003 (\href{https://doi.org/10.1088/1748-0221/15/11/C11003}{2020})

\bibitem{prelPiccolo} %
\footnotesize Benussi, L. \emph{et al.}, \emph{A study of HFO-1234ze (1,3,3,3-Tetrafluoropropene) as an eco-friendly replacement in RPC detectors}, INFN-REPORT-INFN-14-14-LNF (\href{https://arxiv.org/abs/1505.01648}{2015})

\bibitem{ecogas}
\footnotesize Abbrescia, M. \emph{et al.}, \emph{High-rate tests on resistive plate chambers operated with eco-friendly gas mixtures}, EPJ-C 84, 300 (\href{https://doi.org/10.1140/epjc/s10052-024-12545-8}{2024})

\bibitem{ecogasThin}
\footnotesize Abbrescia, M. \emph{et al.}, \emph{Performance of thin-RPC detectors for high rate applications with eco-friendly gas mixtures}, EPJ-C 84, 605 (\href{https://doi.org/10.1140/epjc/s10052-024-12907-2}{2024})

\bibitem{focusPoint} %
\footnotesize Abbrescia, M. \emph{et al.}, \emph{Preliminary results on the long term operation of RPCs with eco-friendly gas mixtures under irradiation at the CERN Gamma Irradiation Facility}, EPJ-Plus 140, 40 (\href{https://doi.org/10.1140/epjp/s13360-024-05773-0}{2025})

\bibitem{pfas1}
\footnotesize Young, C.J. \textit{et al.}, \emph{Atmospheric perfluorinated acid precursors: chemistry, occurrence, and impacts}, Rev. Environ. Contam. Toxicol. 208, 1-109 (\href{https://pubmed.ncbi.nlm.nih.gov/20811862/}{2010})

\bibitem{pfas2}
\footnotesize George, C. \textit{et al.}, \emph{Kinetics of mass transfer of carbonyl fluoride, trifluoroacetyl fluoride, and trifluoroacetyl chloride at the air/water interface}, J. Phys. Chem. 98, 10857–10862 (\href{https://doi.org/10.1021/j100093a029}{1994})

\bibitem{pfas3}
\footnotesize David, L.M. \textit{et al.}, \emph{Trifluoroacetic acid deposition from emissions of HFO-1234yf in India, China and the Middle East}, Atmos. Chem. Phys. 21, 14833–14849 (\href{https://doi.org/10.5194/acp-21-14833-2021}{2021})

\bibitem{GIF++} %
\footnotesize Pfeiffer, D. \emph{et al.}, \emph{The radiation field in the Gamma Irradiation Facility GIF++ at CERN}, NIM A, vol. 866 91-103 (\href{https://doi.org/10.1016/j.nima.2017.05.045}{2017})

\bibitem{AbbrProc}
\footnotesize Abbrescia, M. \emph{et al.}, \emph{Performance and long-term ageing studies on Eco-Friendly Resistive Plate Chamber detectors}, published in these Proceedings

\bibitem{stdCMS}
\footnotesize M.A. Shah \emph{et al.}, \emph{The CMS RPC detector performance and stability during LHC RUN-2}, JINST 14 C11012 (\href{https://iopscience.iop.org/article/10.1088/1748-0221/14/11/C11012}{2019})

\bibitem{stdATLAS}
\footnotesize G.L. Alberghi, \emph{Performance of the ATLAS RPC first level Muon trigger during the Run-2 data takin}, JINST 14 C06007 (\href{https://iopscience.iop.org/article/10.1088/1748-0221/14/06/C06007}{2019})


\end{thebibliography}
\end{document}